# Towards Lateral Inhibition and Collective Perception in Unorganised Non-Neural Systems

Jeff Dale Jones

**Abstract** Lateral Inhibition (LI) phenomena occur in a wide range of neural sensory modalities, but are most famously described in the visual system of humans and other animals. The general mechanism can be summarised as when a stimulated neuron is itself excited and also suppresses the activity of its local neighbours via inhibitory connections. The effect is to generate an increase in contrast between spatial environmental stimuli. Simple organisms, such as the single-celled slime mould *Physarum polycephalum* possess no neural tissue yet, despite this, are known to exhibit complex computational behaviour. Could simple organisms such as slime mould approximate LI without recourse to neural tissue? We describe a model whereby LI can emerge without explicit inhibitory wiring, using only bulk transport effects. We use a multi-agent model of slime mould to reproduce the characteristic edge contrast amplification effects of LI using excitation via attractant based stimuli. We also explore a counterpart behaviour, Lateral Activation (where stimulated regions are inhibited and lateral regions are excited), using simulated exposure to light irradiation. In both cases restoration of baseline activity occurs when the stimuli are removed. In addition to the enhancement of local edge contrast the long-term change in population density distribution corresponds to a collective response to the global brightness of 2D image stimuli, including the scalloped intensity profile of the Chevreul staircase and the perceived difference of two identically bright patches in the Simultaneous Brightness Contrast (SBC) effect. This simple model approximates LI contrast enhancement phenomena and global brightness perception in collective unorganised systems without fixed neural architectures. This may encourage further research into unorganised analogues of neural processes in simple organisms and suggests novel mechanisms to generate collective perception of contrast and brightness in distributed computing and robotic devices.

Jeff Dale Jones
Centre for Unconventional Computing, University of the West of England, Bristol, UK, e-mail: jeff.jones@uwe.ac.uk

2    Jeff Dale Jones# 1 Introduction

Living organisms perceive their environment with a wide variety of special sensory modalities. Enhancing the contrast in the stream of information from these senses allows organisms to discriminate between small changes in signal level, potentially enhancing survivability. Lateral Inhibition (LI) is a neural mechanism which enhances the activity of neurons directly exposed to excitatory stimuli whilst suppressing the activity of their near neighbours (see Fig. 1 for a schematic illustration). LI phenomena have been described in auditory [9], somatosensory [22] and olfactory senses [29], but are most famously described in the visual systems of a wide range of animals, including humans [8, 15, 14].

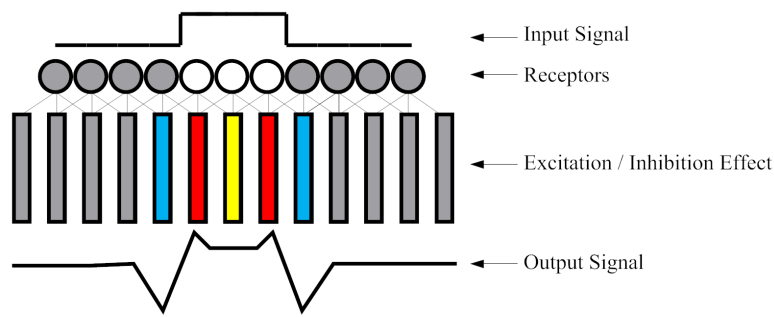

**Fig. 1** Schematic illustration of Lateral Inhibition response. Original stimulus (top) activates neurons (circled) whose local inhibitory wiring results in excitation and inhibition effects in neighbouring neurons (bars), resulting in the enhanced output signal (bottom).

# 2 Non-neural Computation in Slime Mould

LI phenomena, and the mechanisms which generate them, result in an effective and efficient means of enhancing environmental perception in organisms containing nervous systems ranging from the most complex to the most primitive. But can such mechanisms occur in organisms which do not possess any nervous system? The true slime mould *Physarum polycephalum* is a single-celled amoeboid organism with a very complex life cycle. The plasmodium stage, where a giant syncytium formed by repeated nuclear division is encompassed within a single membrane, has been shown to exhibit a complex range of biological and computational behaviours. The plasmodium of *P. polycephalum* is comprised of a complex gel/sol transport network which is continuously remodelled in response to its environment. The organism behaves as a distributed computing *material*, capable of responding to a wide range of spatially represented stimuli. The ectoplasmic gel phase is composed of a sponge-like matrix of contractile actin and myosin fibres through which the endoplasmic sol flows. Local oscillations in the thickness of the plasmodium spontaneously appear with approximately 2 minutes duration [24]. The spatial and temporal organisation of the oscillations has been shown to be extremely complex [25] and affects the internal movement of sol through the network by local assembly and disassembly of the actin-myosin structures. The protoplasm moves backwards and forwards within the plasmodium in a characteristic manner known as shuttle-streaming.

The plasmodium is able to sense local concentration gradients and the presence of nutrient gradients appears to alter the structure of external membrane areas. The softening of the outer membrane causes a flux of protoplasm towards the general direction of the gradient in response to internal pressure changes caused by the local thickness oscillations. The strong coupling between membrane contraction and streaming movement is caused by the incompressibility of the fluid requiring a constant volume - the weakening of the membrane provides an outlet for the pressure. When the plasmodium has located and engulfed nearby food sources the material behind the solid active growth front begins to form a network of protoplasmic veins which become coarser in distal regions. These veins connect nutrient sources in networks that have been shown to be efficient in terms of their overall distance and resilience to random damage [16] (see Fig. 2 for an example transport network connecting oat flake nutrient sources).

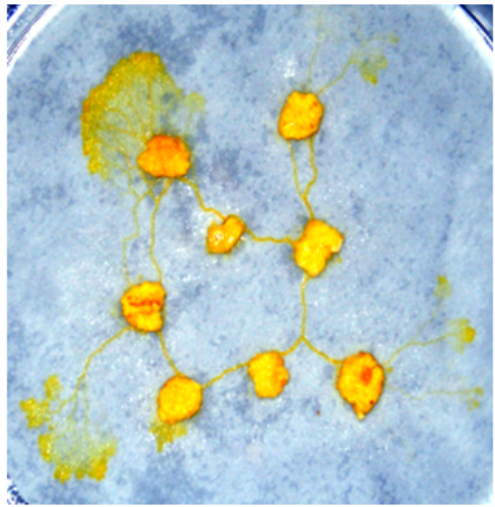

**Fig. 2** Plasmodium of slime mould *Physarum polycephalum* inoculated on damp filter paper forms a transport network connecting oat flake nutrients. Note the fan-shaped solid active growth front of the plasmodium at the top-left of the image, with the network forming and coarsening behind the active front.

The transport network distributes protoplasm (including microscopic nutrients) throughout the organism. The relative simplicity of the cell and the distributed nature of its control system make *P. polycephalum* a suitable subject for research into distributed computation substrates. In recent years there have been a large number of studies investigating its computational abilities, prompted by Nakagaki et al. who reported the ability of *P. polycephalum* to solve path planning problems [17]. Subsequent research confirmed and broadened the range of abilities to spatial representations of various graph problems [16, 23, 2], combinatorial optimisation problems [5], construction of logic gates [27] and logical machines [1], [4], and as a means to achieve distributed robotic control [28], robotic manipulation [3] and robotic amoeboid movement [10], [13].

In this paper we take inspiration from very simple organisms, such as slime mould, without neural tissue which nevertheless exhibit very complex behaviours. Although *Physarum* does not possess neural tissue, it displays behaviour which is analogous to neural spiking and synaptic learning [7], [19]. *Physarum* has been used to study potential low-level mechanisms of perception in the Kanizsa illusory contour phenomena [26] and other simple collective organisms have been used to study geometric illusions, such as the Muller-Lyer effect [21]. In this paper we initially start from a simpler level, describe collective mechanisms by which sensory contrast enhancement phenomena analogous to Lateral Inhibition phenomena can emerge in disorganised non-neural systems.

We use a multi-agent particle based model of slime mould to demonstrate and elucidate the low-level behaviours which generate these collective phenomena. In Section 3 we describe the multi-agent model of slime mould, its previous use as a material computing substrate, and parameters governing its use for the experiments in this chapter. In Section 4 we examine the emergence of LI phenomena in the model in response to presentation with attractant stimuli. The opposite response — Lateral Activation — is described in response to presentation with adverse stimuli (simulated light irradiation) in Section 5. In Section 6 we build on these simple mechanisms by showing how a *global* collective representation of an environment (in this case an approximation of overall brightness) can be generated by bulk drift of population density in response to greyscale spatial patterns presented as attractant stimuli. We conclude in Section 7 by summarising the results, the main contributions of this chapter, and examining potential applications of these collective phenomena for computing and robotics applications.

## 3 Multi-Agent Model of Slime Mould

We used a multi-agent approach to generate the *Physarum*-like behaviour. This approach was chosen specifically specifically because we wanted to reproduce the generation of complex behaviour in slime mould using the same limitations that slime mould has, i.e. using very simple component parts and interactions, and no special or critical component parts to generate the emergent behaviour. Although other modelling approaches, notably cellular automata, also share these properties, the direct mobile behaviour of the agent particles renders it more suitable to reproduce the flux within the *Physarum* plasmodium. The multi-agent particle model [12] uses a population of coupled mobile particles with very simple behaviours, residing within a 2D diffusive lattice. The lattice stores particle positions and the concentration of a local diffusive factor referred to generically as chemoattractant. Particles deposit this chemoattractant factor when they move and also sense the local concentration of the chemoattractant during the sensory stage of the particle algorithm. Collective particle positions represent the global pattern of the material. The model runs within a multi-agent framework running on a Windows PC system. Performance is thus influenced by the speed of the PC running the framework. The particles act independently and iteration of the particle population is performed randomly to avoid any artifacts from sequential ordering.

## 3.1 Generation of Collective Cohesion Phenomena

The behaviour of individual particles occurs in two distinct stages, the sensory stage and the motor stage. In the sensory stage, the particles sample their local environment using three forward biased sensors whose angle from the forwards position (the sensor angle parameter, SA), and distance (sensor offset, SO) may be parametrically adjusted (Fig. 3a). The offset sensors generate local indirect coupling of sensory inputs and movement to generate the cohesion of the material. The SO distance is measured in pixels and a minimum distance of 3 pixels is required for strong local coupling to occur. It was shown in [11] that large SO values result in regular self-organised domains ('vacancy islands') of small vacant regions in the material. This would not be desirable in these experiments as we gauge the response of the material by measuring population distribution density (which would be affected by the presence of these domains), so we randomly selected the SO parameter for each particle, at each scheduler step, from the range of 1 – 20 pixels. This maintained the cohesion of the material whilst avoiding the formation of these vacancy domains. During the sensory stage each particle changes its orientation to rotate (via the parameter rotation angle, RA) towards the strongest local source of chemoattractant (Fig. 3b). Variations in both SA and RA parameters have been shown to generate a wide range of reaction-diffusion patterns [11] and for these experiments we used SA 60 and RA 60. After the sensory stage, each particle executes the motor stage and attempts to move forwards in its current orientation (an angle from 0–360 degrees) by a single pixel forwards. Each lattice site may only store a single particle and particles deposit chemoattractant into the lattice (5 units per step) only in the event of a successful forwards movement. If the next chosen site is already occupied by another particle move is abandoned and the particle selects a new randomly chosen direction.

## 3.2 Representation of Spatial Stimuli

The presentation of simple uniform attractant stimuli to the virtual plasmodium was achieved by incrementing attractant values by 1.275 units at stimulated regions every scheduler step. The presentation of Adverse stimuli (simulated light irradiation) to the model was achieved by reducing the sensitivity of the particle sensors in illuminated regions by 80% and reducing particle chemoattractant deposition in the same regions by 80%.

Representation of more complex stimuli with difference brightness levels (Section 6) was achieved by incrementing attractant values by the corresponding stimuli image pixel brightness at each location on the lattice and scaling this value downwards by multiplying by 0.01. The reduction in attractant concentration by scaling reduces the attractant stimuli concentration towards the baseline flux generated by the particle movement and maintains the integrity of the virtual plasmodium (stronger stimuli would cause the material to tear). Diffusion within in the lattice was implemented at each scheduler step and at every site in the lattice via a simple mean filter of kernel size $5 \times 5$. Damping of the diffusion distance, which limits the distance of chemoattractant gradient diffusion, was achieved by multiplying the mean kernel value by 0.95 per scheduler step.

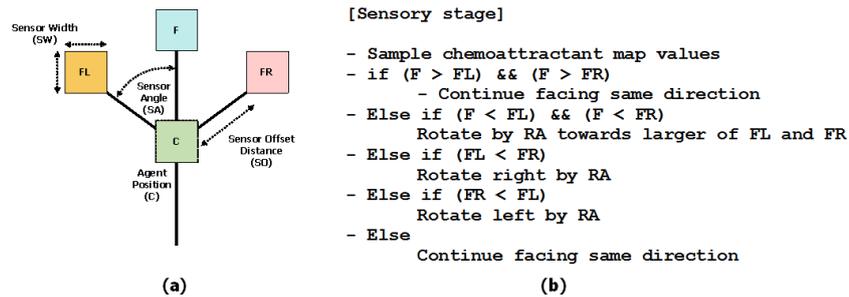

**Fig. 3** Architecture of a single particle of the virtual material and its sensory algorithm. (a) Morphology showing agent position 'C' and offset sensor positions (FL, F, FR), (b) Algorithm for particle sensory stage.

## 3.3 Translation of Neuronal Responses to Non-neural Mechanisms

To explore mechanisms corresponding to LI in non-neural systems we must translate the relevant neuroscience terminology into terms which can be represented in the model system. Input stimuli (such as light to the human visual system) can be represented in the model by spatial projection of simulated chemoattractants. These stimuli attract the particles comprising the virtual plasmodium. We may say that the projection of chemoattractants results in an excitatory response. Conversely we can generate an inhibitory response by the spatial projection of repellents or other adverse stimuli. Slime mould is known to avoid illumination with certain wavelengths of visible light and we can use this feature to generate an inhibitory response in the model.

The basic responses of excitation (by chemoattraction) and inhibition (by repulsion or irradiation) are short term approximations of the neural response. To approximate long-term neural responses (changes in spatio-temporal patterns of neural activity) we must utilise the changing spatial density distribution of the particles comprising the model material, in response to the projected stimuli, i.e. changes in the population density correspond to regions of increased / decreased neural activity. The model must also respond to the withdrawal of stimuli. In neural systems this would result in a reversion to baseline activity. In the model this must be represented by the restoration of uniform population density on withdrawal of the stimuli.

## 4 Lateral Inhibition Phenomena Using Attractant Stimuli

We initialised the virtual plasmodium comprising 8000 particles within a $300 \times 100$ pixel tube-like horizontal arena bordered by inhabitable areas on the top and bottom and open ended left and right edges. Periodic boundary conditions were enforced. We measured population density across the arena by counting the number of particles in the Y-axis for each X-axis position. We recorded population density every 10 scheduler steps. After initialisation the population, constrained by the architecture of the arena, formed a single tube with relatively uniform population density (Fig. 4,a).

An attractant stimulus was presented to the virtual plasmodium after 500 scheduler steps by projecting chemoattractant into the middle-third habitable section of the arena (white region in Fig. 4,c). The attractant stimulus caused increased flux of particles into the stimulus area, an increase in population density in this area, and a corresponding decrease in density outside the stimulus region (Fig. 4,e). Upon removal of the stimulus after 4000 steps the population was no longer attracted to the central region and the tube adapted its shape in response to the uniform chemoattractant profile (Fig. 4,g). The population density eventually returned to uniform density across the arena (Fig. 4,i).

A space-time plot of the population density indicates how the changes in population density are initiated at the stimulus boundaries and propagate outwards from these regions (Fig. 5). Regions inside the attractant stimulus area correspond to excitation areas and regions outside correspond to inhibited activity. The presentation of attractant stimuli, and the response of the particle population, causes an increase in signal contrast (measured in terms of population density) between stimulated and non-stimulated areas (Fig. 6).

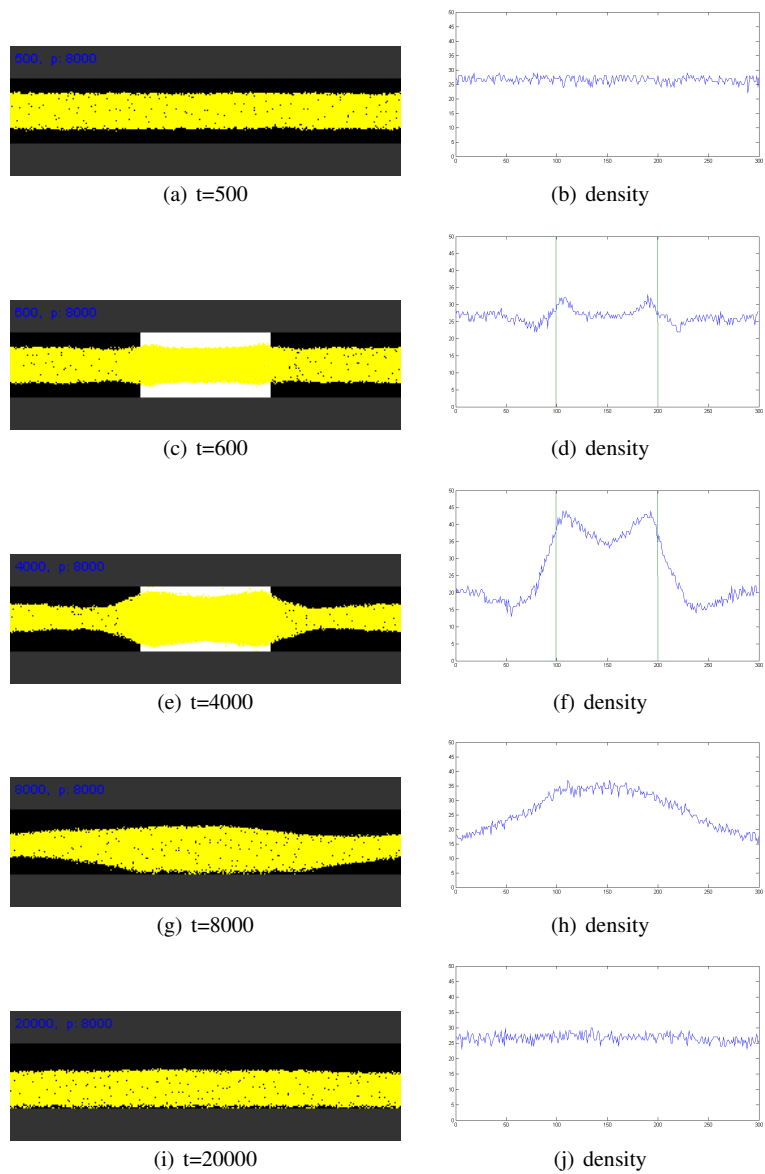

**Fig. 4** Response of virtual plasmodium flux to attractant stimulus. a) population initialised within horizontal arena forms single tube, c) presentation of attractant stimulus bar (light area) results in flux towards stimulus area, e) population density is increased at stimulus region and reduced at unstimulated region, g) removal of stimulus results in adaptation to uniform attractant profile, i) uniform density is restored, b,d,f,h,j) cross-section plots of population density across the arena.

Lateral Inhibition and Collective Perception in Unorganised Non-Neural Systems  9

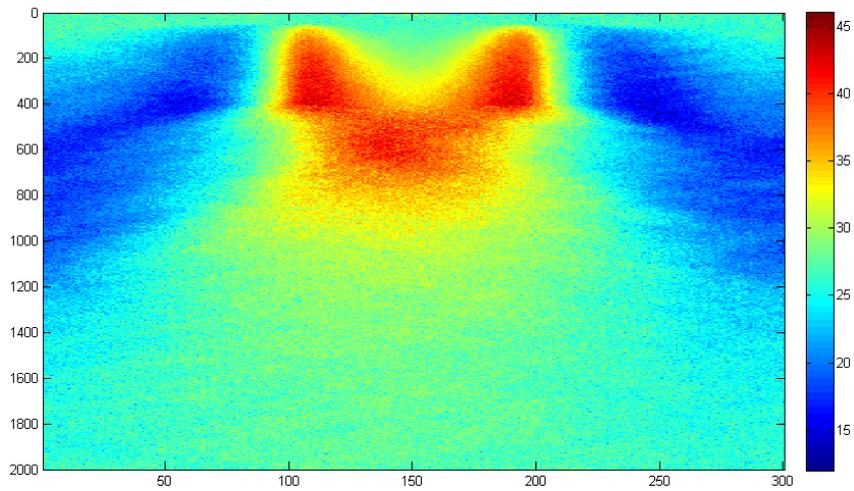

**Fig. 5** Space-time plot of population density flux under attractant stimuli conditions, time proceeds downwards. Note that changes in density are initiated at borders of the stimulus boundary.

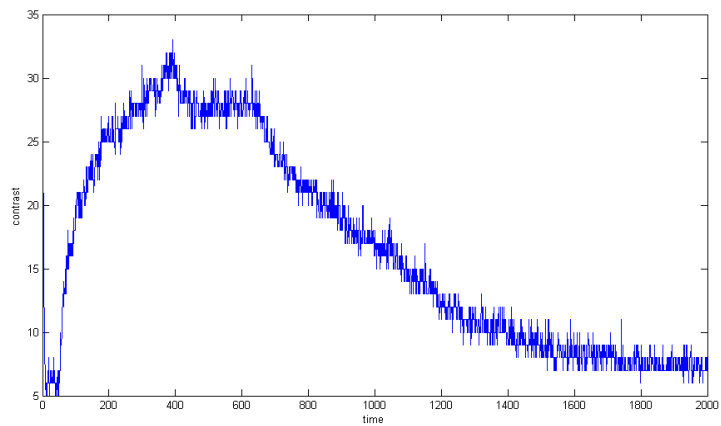

**Fig. 6** Presentation of attractant stimuli and its effect on population density, interpreted as signal contrast enhancement. Stimulus presented after 50 samples (t=500), stimulus removed after 400 samples (t=4000) causing gradual reversion to baseline activity.

## 5 Lateral Activation Phenomena Using Adverse Stimuli

To examine the collective response to Adverse Stimuli (simulated exposure to illumination, which the *Physarum* plasmodium avoids), we used the same arena with the virtual plasmodium inoculated as a horizontal strip with approximately uniform density (Fig. 7, a and density profile in b). The stimulus pattern was again in the central third of the arena, but the attractant region was replaced with a region of simulated illumination. Particles at the border of exposed areas preferentially moved to unexposed regions and the local coupling of particles resulted in collective flux away from the stimulus area (Fig. 7, c and e). The resulting density profile (Fig. 7, d and f) demonstrates the inhibition effect within the illuminated region whilst the unexposed neighbouring regions show an increase in population density. When the adverse stimulus was removed from the central region the population density re-normalised to a uniform level within 15000 steps (Fig. 7, g and i and corresponding density profiles in h and j respectively). The space-time plot of changing population density shows the inhibition effect in the central region and the lateral propagation of increased density (Fig. 8). Areas inside the adverse stimulus correspond to an inhibition response and areas outside correspond to excited activity. As in the attractant stimuli case, the presentation of adverse stimuli increases the signal contrast (population density) between illuminated and non-illuminated regions (Fig. 9).

## 6 From Low-level Mechanisms to Unorganised Collective Perception

The changes in population density over time in response to patterns of attractant and adverse stimuli correspond to unorganised Lateral Inhibition and Lateral Activation mechanisms respectively. How do these mechanisms respond to more complex arrangements of stimuli? We examine the response of the virtual plasmodium to attractant stimuli in the pattern of the Chevreul staircase illusion (Fig. 10a). The Chevreul staircase is a sequence of identical width uniform vertical bars. Each bar is lighter in intensity then its leftmost neighbour (see cross-section intensity plot in Fig. 10e). Although each bar is uniform in intensity the image is typically perceived as having a scalloped profile across each bar, i.e. the left side of each bar (when adjacent to a darker bar) is perceived as being lighter, and the right side of each bar (when adjacent to a lighter bar) is perceived as being darker. The mechanisms underlying this illusory percept are considered to be mediated by activity at the retinal and cortical levels [18], [20].

We initialised the virtual plasmodium (comprising 169,402 particles) on an arena patterned with the Chevreul staircase image (692 × 288 pixels). Each vertical bar corresponded to increasing concentration stimuli. Periodic boundary conditions were used and two bordering bars offering no stimuli were placed at the left and right of the image. The initially uniform distribution of particles (Fig. 10b) was affected by the attractant stimuli and particles migrated to regions of higher concentration. The flux of particles changed the population density, with higher occupancy emerging in regions which corresponded to lighter areas of the original image (Fig. 10c). Note that within each bar the density is greater towards the left side of the bar. This is caused by influx from the darker bar to the left and efflux towards lighter bars to the right. A plot of the population density at 400 scheduler steps demonstrates both the scalloped response to each bar junction and the global increase in density towards the lighter bars (corresponding to the perceived lightness of the global image, Fig. 10e). A space-time plot of the evolution of population density indicates that the scalloped effect occurs immediately after presentation with the stimuli (Fig. 11). The differences in density between the bars also starts immediately after presentation but the full increase in 'brightness' between each bar occurs much later, due to the time taken for particles to migrate towards higher concentration areas (Fig. 10e, note the gradual increase in density in each bar over time). The increase in contrast of the global (entire arena) population density over time can be seen in Fig. 12. This contrast is caused by the relatively slow flux of particles across the entire arena and corresponds to a coarse representation of the image brightness.

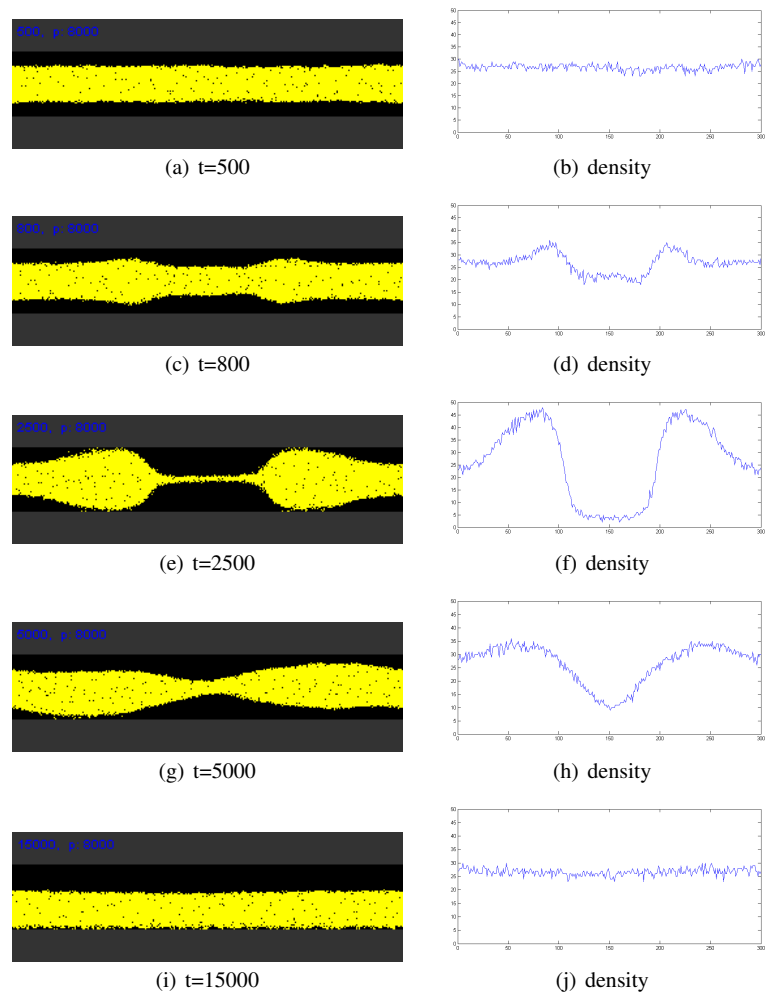

**Fig. 7** Response of virtual plasmodium flux to simulated light irradiation. a) population initialised within horizontal arena forms single tube, c) presentation of simulated light irradiation (centre, not shown) results in flux away from irradiated area, e) population density is decreased at irradiated region and increased at unexposed region, g) removal of adverse stimulus results in increased flux to inner region, i) uniform density is restored, b,d,f,h,j) cross-section plots of population density across the arena.

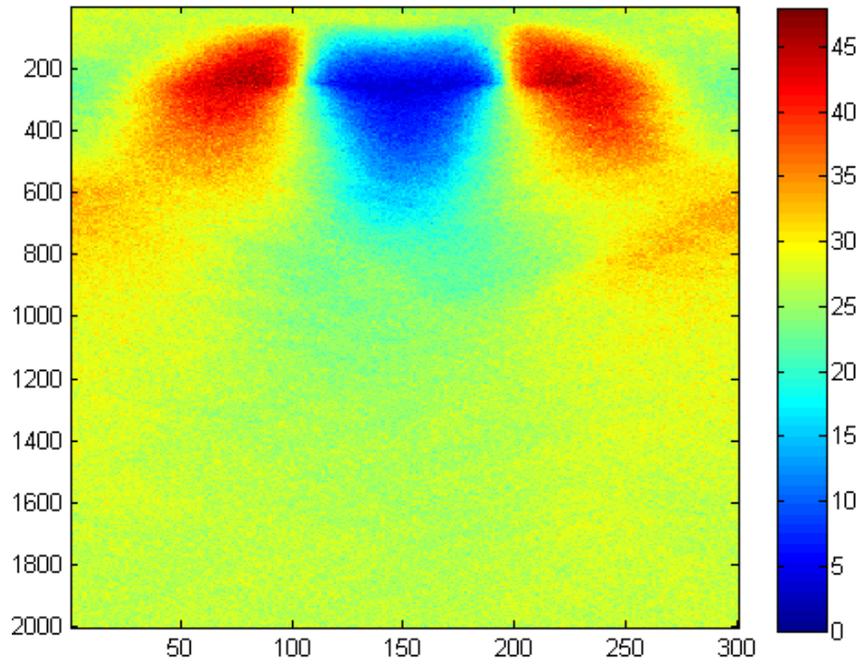

**Fig. 8** Space-time plot of population density flux under adverse stimuli condition, time proceeds downwards. Note that changes in density are initiated at borders of stimulus boundary.

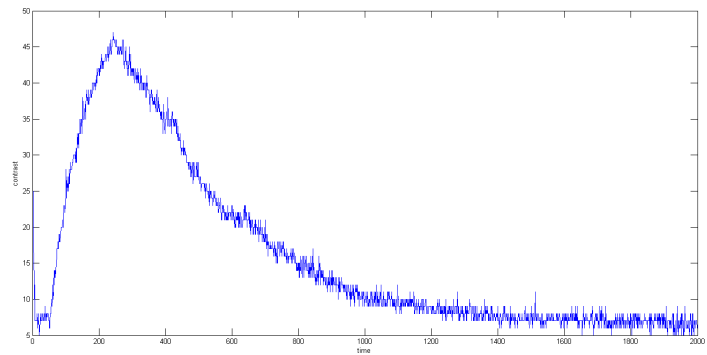

**Fig. 9** Presentation of adverse stimuli (simulated light irradiation) and its effect on population density, interpreted as signal contrast enhancement. Adverse stimulus presented after 50 samples (t=500), stimulus removed after 250 samples (t=2500) causing gradual reversion to baseline activity.

The Chevreul staircase consists of contiguous regions of gradually increasing lightness. Another illusory perception of lightness occurs in regions with non-contiguous increases in lightness. This is known as the Simultaneous Brightness Contrast (SBC) effect.[1]

A simple example of SBC is shown in Fig. 13a. The image consists of two large squares, the left-most square in dark grey and the right-most square in a lighter grey. Overlaying the centre of each square is a vertical band of grey (intermediate in lightness between the left and right squares). Although the vertical grey bands are of equal lightness (see cross-section in 13d), their lightness is perceived differently: the band on the left is typically perceived as lighter than the band on the right. As with the Chevreul staircase the explanatory mechanisms for this illusion have been suggested as LI at the retinal and cortical level. How does the virtual plasmodium respond when presented with this stimuli?

We initialised the model on a $600 \times 300$ lattice with 153,000 particles (approximately the same density as used in the Chevreul staircase experiments) with uniform initial distribution (13b) and periodic boundary conditions. The lightness differences of the SBC image areas were represented as differing attractant concentration profiles. The population response is to migrate from areas of low concentration (brightness) towards areas of high concentration. This flux is mobilised at the junctions between light and dark regions and a global difference in population density is produced (13c). Because the vertical grey band in the darker square is surrounded by a darker region, it receives influx from the darker square. The vertical band within the lighter square is surrounded by a lighter region and there is an efflux of particles to the lighter square. Simultaneously there is transport of particles at the centre of the image (the border between the two large squares) towards the lighter side, and also

---

[1] note that the terms *lightness* (SLC) and *brightness* (SBC) are both used in the literature to describe the same perceptual effect, though their actual definitions are somewhat different, depending on the particular presentation of the stimuli.

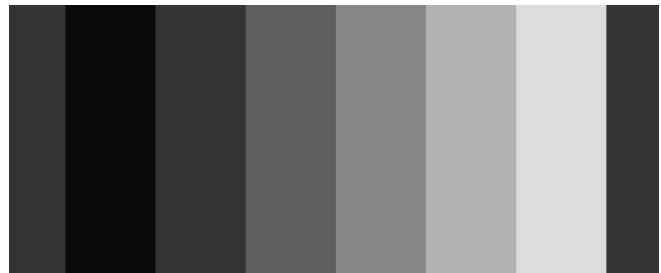

(a) Chevreul Staircase

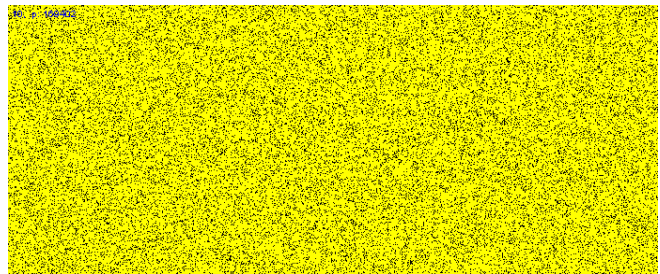

(b) t=10

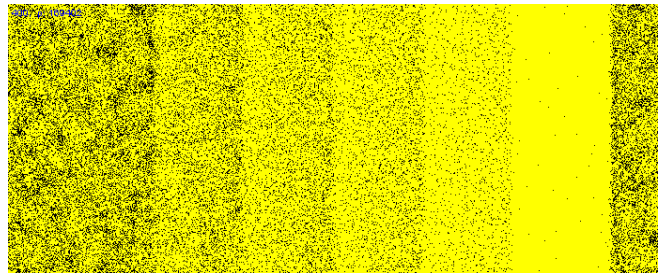

(c) t=400

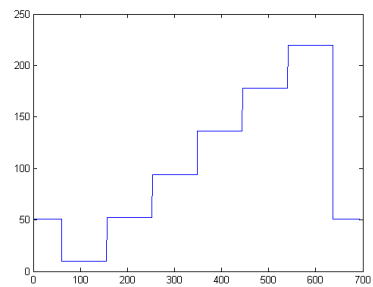 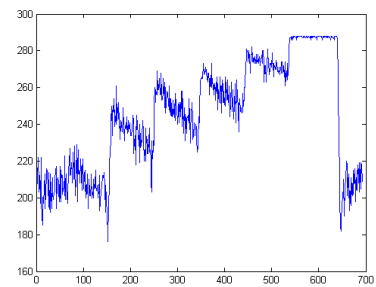

(d) original section    (e) population density t=400

**Fig. 10** Collective representation of the Chevreul staircase illusion. a) Original greyscale Image of uniform bars of increasing lightness (surrounded by left and right borders), b) initial uniform population distribution of particle population, c) population distribution after 4000 scheduler steps showing increased population density at lighter (greater attractant concentration) regions, d) cross-section profile of original image stimulus, e) cross-section of population density at t=400 showing scalloped borders and increasing contrast.

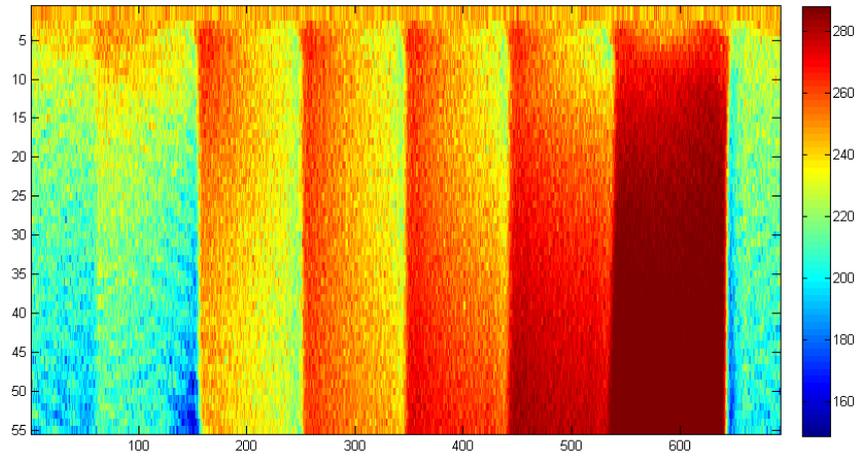

**Fig. 11** Space-time plot of population density flux under Chevreul illusory stimulus, time proceeds downwards.

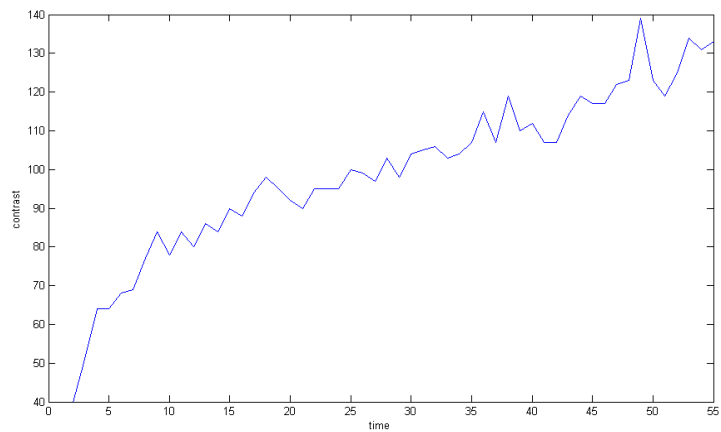

**Fig. 12** Plot of the emergence of global contrast by particle flux in response to the Chevreul staircase stimulus (difference in population density range, from an initially random density distribution, to halting at 550 scheduler steps).

at the left and right edges of the image (due to periodic boundary conditions). The resulting population density plot (13e) shows that the vertical band in the left square is perceived as 'brighter' than the band in the right square (although the stimuli concentration presented was identical). This illusory percept matches that of human subjects when presented with the same stimulus type [6].

A space-time plot of the evolution of population density demonstrates the differences in population density emerging as the experiment progresses. As with the Chevreul staircase, the local response to adjacent stimuli in the SBC image is instantaneous, whereas the global response to overall brightness changes in the image takes significantly longer, (also seen in the contrast plot in Fig. 15), again caused by the relatively slow movement of particles across the image areas.

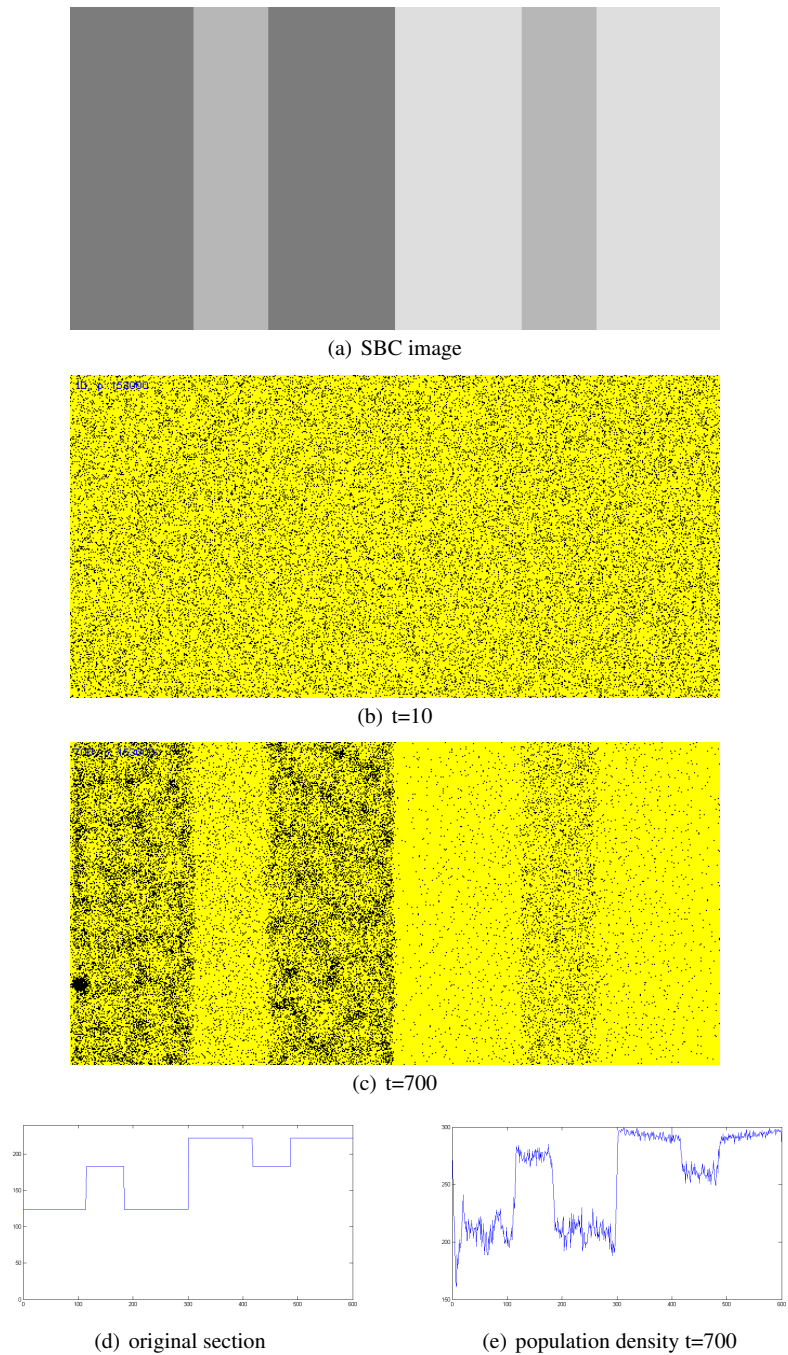

(a) SBC image

(b) t=10

(c) t=700

(d) original section

(e) population density t=700

**Fig. 13** Collective representation of the Simultaneous Brightness Contrast (SBC) illusion. a) Original greyscale Image of two large squares, each with a central band of identically light grey, b) initial uniform population distribution of particle population, c) population distribution after 700 scheduler steps showing increased population density at lighter (greater attractant concentration) regions and greater density in the left central grey strip, d) cross-section profile of original image stimulus, e) cross-section of population density at t=700 showing illusory percept of the left central grey strip as brighter than the right central grey strip.

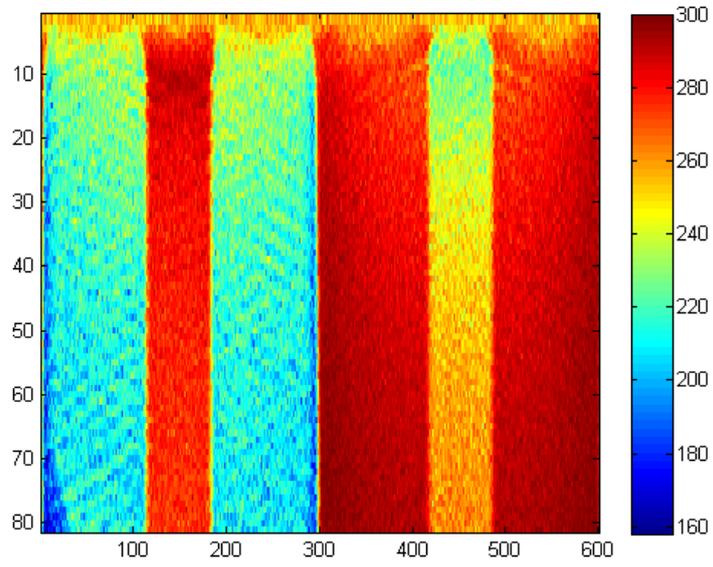

**Fig. 14** Space-time plot of population density flux under Simultaneous Brightness Contrast stimulus, time proceeds downwards.

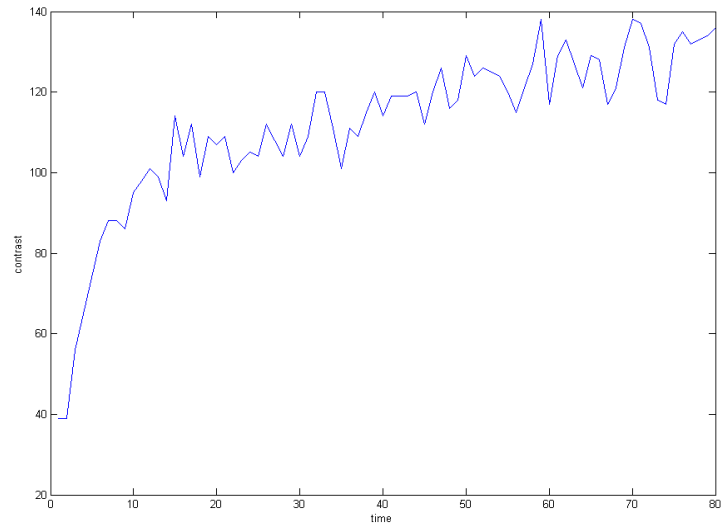

**Fig. 15** Plot of the emergence of global contrast over time by particle flux in response to the SBC stimulus (difference in population density range, from an initially random density distribution, to halting at 800 scheduler steps).

# 7 Conclusions

We have demonstrated the results of modelling experiments into the generation of spatial contrast enhancement analogous to Lateral Inhibition in unorganised non-neural systems using a multi-agent model of slime mould *Physarum polycephalum*. The results show the classic LI contrast enhancement in response to attractant stimuli and its opposite counterpart behaviour (Lateral Activation) in response to adverse stimuli (simulated light irradiation). These effects does not require pre-existing inhibitory connectivity and are generated by bulk transport of the particles comprising the virtual material, initiated at the borders of stimuli projection. Restoration of uniform baseline activity (population density distribution) is established when stimuli are removed. In addition to the local edge contrast enhancement we observed long-term changes in population density distribution when the population was presented with a more complex attractant stimulus pattern. This was caused by the flux of particles towards brighter regions of the image and corresponds to a (rather crude) collective response to the global brightness of the original image stimuli, including the scalloped intensity profile of the Chevreul staircase and the perceived difference of two identically bright patches in the Simultaneous Brightness Contrast (SBC) effect. Interestingly, this simple mechanism reproduces the illusory human percepts in both the Chevreul staircase and SBC figures.

How may these simple mechanisms correspond to real-life perception in neural systems? It is possible to relate the response of the model to attractant and repellent stimuli to the response of different populations (ON or OFF type respectively) of bipolar cells in the retina of the human visual system. In the model, however, we only have a single 'type' of particle and the opposing response to attractant and repellent stimuli loosely correspond to the response to light and dark stimuli in the retina. The receptive field around retinal stimuli is approximated by the region of influx of efflux of particles from the local stimuli and the representation of the *eigengrau* level (the baseline activity in the absence of any stimulus) may be approximated by the population density of the particles when no attractant stimuli or repellent stimuli are present.

The most notable feature of this approach is that LI phenomena can be approximated without explicit fixed inhibitory connections. This suggests possible mechanisms by which simple organisms without neural tissue may achieve sensory contrast enhancement. In organisms such as slime mould the internal protoplasmic transport of cellular material could be harnessed to generate the LI mechanism. The sensory contrast enhancement afforded by LI could allow for the enhancement of weak spatial stimuli (such as nutrient location). Conversely, the LA mechanism would amplify weak hazardous stimuli, providing alternate migration paths away from hazardous regions. The relatively slow restoration of baseline activity also allows a temporary memory effect denoting the approximate location of attractant and hazard stimuli. Because the LI and LA phenomena in this model do not rely on fixed inhibitory connectivity it is particularly suited to systems and organisms which have adaptive architectures and body plans respectively.

In the context of adaptive materials and robotics applications the mechanism illustrates how complex sensory behaviour can be distributed within an unorganised material (or robotic collective) itself. This allows greater freedom from having to pre-specify connectivity to implement sensory contrast enhancement and allows redundancy for individual faulty components. We hope that ongoing research may lead to other unorganised material approximations of complex neural functions seen in brightness perception (including illusory phenomena such as neon colour spreading, illusory contours and brightness assimilation effects), and implementation of other spatial feature detectors (including orientation detection, edge completion, gestalt phenomena, optic flow), and direction discrimination.

## Acknowledgements

This work was supported by the EU research project "Physarum Chip: Growing Computers from Slime Mould" (FP7 ICT Ref 316366)